\documentclass[10pt,twocolumn]{IEEEtran}

\usepackage{amsmath,graphics,amssymb,epsfig,subfigure,color,cite}

\usepackage{array}
\usepackage{multirow}
\usepackage{pifont}
\usepackage{arydshln}
\usepackage{xcolor}
\usepackage{enumerate}
\usepackage{bm}
\usepackage{float}
\usepackage{makecell}
\usepackage[right=0.625in, left=0.625in, top=0.7in, bottom=1.1in]{geometry}

\usepackage{lipsum}
\usepackage{capt-of}

\usepackage{array}
\newcolumntype{C}[1]{>{\centering\arraybackslash}p{#1}}
\newcolumntype{L}[1]{>{\raggedright\arraybackslash}p{#1}}

\usepackage{algorithm}
\usepackage{algpseudocode}
\usepackage{algcompatible}
\algblockdefx{WHILE}{ENDWHILE}[1]%
  {\textbf{while }#1 \textbf{}}%
  {\textbf{end}}
\algblockdefx{FORP}{ENDFORP}[1]%
  {\textbf{for }#1 \textbf{do in parallel}}%
  {\textbf{end}}
\algblockdefx{FOR}{ENDFOR}[1]%
  {\textbf{for }#1 \textbf{do}}%
  {\textbf{end}}
\algblockdefx{IF}{ENDIF}[1]%
  {\textbf{if }#1 \textbf{}}%
  {\textbf{end}} 
\algblockdefx{ELSIF}{ENDIF}[1]%
  {\textbf{else if }#1 }%
  {\textbf{end}}
\algnewcommand\algorithmicprocedure{\textbf{function}}
\algnewcommand\FUNC{\item[\algorithmicprocedure]}%
\algnewcommand\algorithmicendprocedure{\textbf{end function}}
\algnewcommand\ENDFUNC{\item[\algorithmicendprocedure]}%
 \usepackage{setspace}
\let\Algorithm\algorithm
\renewcommand\algorithm[1][]{\Algorithm[#1]\setstretch{1.4}}

\usepackage{bbm}
\usepackage{bigints}
\usepackage{adjustbox}

\usepackage{hhline}

\floatname{algorithm}{Algorithm}

\usepackage{enumitem}

\makeatletter
\newcommand{\vast}{\bBigg@{4.5}}
\newcommand{\Vast}{\bBigg@{7.5}}
\makeatother

\usepackage{tabularray}
\usepackage{xcolor}

\definecolor{rowA}{RGB}{255,255,255}
\definecolor{rowB}{RGB}{255,255,255}
\definecolor{rowC}{RGB}{255,255,255}
\definecolor{headercolor}{RGB}{245,245,245}

\begin{document}
\title{\fontsize{21}{28}\selectfont Rethinking the Foundations of Two-Sided AI Models for 6G
}

\author{Yongjeong Oh, Zihan Chen, Timothy J. O’Shea, Junyong Shin, Jinho Choi, Yo-Seb Jeon, and Jihong Park
        \thanks{Y. Oh, Z. Chen, and J. Park are with Singapore University of Technology and Design, Singapore 487372 (email: \{yongjeong\_oh, zihan\_chen, jihong\_park\}@sutd.edu.sg).} 
        \thanks{T. J. O’Shea is with DeepSig Inc., Arlington, VA 22203, USA (email: tim@deepsig.ai)} 
        \thanks{J. Shin and Y.-S. Jeon are with POSTECH, Pohang, Gyeongbuk 37673, Republic of Korea (email: \{sjyong, yoseb.jeon\}@postech.ac.kr).} 
        \thanks{J. Choi is with the University of Adelaide, SA 5005, Australia (email: jinho.choi@adelaide.edu.au).} 
        \thanks{(Corresponding authors: J. Park, Y.-S. Jeon).}
        \vspace{1mm}}
	\vspace{-2mm}

\maketitle

\begin{abstract} 
For next-generation air interfaces, two-sided artificial intelligence (AI) models have received growing attention, with AI models deployed at both the transmitter and receiver for efficient channel feedback and data communication. However, their practical deployment is complicated by assumptions commonly made in existing studies, including isolation from legacy users, training under predefined channel conditions, and gradient-based fine-tuning requiring substantial cross-vendor communication. This article revisits these assumptions and presents practical alternatives. First, for legacy coexistence, we integrate two-sided model processing into the 5G New Radio (NR) protocol stack and validate its operation alongside conventional NR on a real-world testbed. 
Second, instead of training under a massive number of predefined channel conditions, we construct a compact model table by jointly optimizing two-sided models with trainable surrogate channels, and select the best model according to the current channel condition to enable channel adaptation with high task performance and low training/storage overhead. 
Finally, unlike existing fine-tuning that exchanges large gradient vectors containing potentially private model information, we present gradient-free zeroth-order fine-tuning that requires only scalar feedback, facilitating multi-vendor interoperability. Together, these approaches advance the practical deployment of two-sided AI models while highlighting key open challenges.
\end{abstract}


\section{Introduction}\label{Sec:Intro}

As wireless networks advance toward more intelligent and adaptive air interfaces, artificial intelligence (AI) is being integrated ever more deeply into both the transmitter and the receiver \cite{11320975}. This trend has spurred several research directions, including neural transceivers, deep joint source-channel coding, and task-oriented semantic communication \cite{11316151,DeepJSCC_Q}. 
Although these directions differ in their objectives and system settings, many rely on a common architecture in which transmitter-side and receiver-side AI models are designed to operate jointly. The 3rd Generation Partnership Project (3GPP) refers to such paired models as \textit{two-sided AI models}, particularly when the models are deployed across the user equipment (UE) and the next-generation Node B (gNB) \cite{3gpp_tr38843_rel19}. 

While early academic research on two-sided models primarily focused on demonstrating performance gains under controlled training and evaluation settings, recent industry and standardization efforts have begun exploring their use for practical use cases such as channel state information (CSI) feedback \cite{3gpp_tr38843_rel19,9970357}. In translating these academic advances into real-world deployments, such efforts have naturally adopted the methodological foundations established in the literature. These foundations typically assume that a two-sided model operates in isolation from legacy UEs, that training is performed under predefined channel conditions, and that the transmitter and receiver models are fine-tuned through backpropagation. 
Although these standard assumptions simplify model development and help isolate algorithmic gains, they merely reflect specific design choices rather than intrinsic or necessary properties of two-sided models.

When carried into real-world deployments, these inherited foundations give rise to significant practical challenges. Isolation from legacy UEs does not account for coexistence with conventional radio access technologies (RATs). Training under prescribed channel conditions creates a trade-off between specialization and generalization in channel adaptation, as maintaining channel-specific models increases training and storage costs, whereas a single model trained across diverse conditions compromises site-specific performance. Gradient-based fine-tuning becomes difficult when the UE-side and gNB-side models are developed by different vendors, as exchanging model parameters or gradients risks exposing proprietary model information while adding communication and memory overhead. 

These observations motivate a fundamental rethinking of the foundations of two-sided AI models. From this perspective, this article revisits the prevailing assumptions and discusses practical alternatives for coexistence, channel adaptation, and model updates. Specifically, we examine (i) how two-sided models can coexist with legacy 5G New Radio (NR) systems, (ii) how they can adapt to diverse channel conditions without relying on channel-specific training, and (iii) how transmitter-side and receiver-side models can be updated across vendors without gradient backpropagation. Finally, we discuss open research challenges and future directions toward practical two-sided AI models for next-generation wireless networks.

\begin{figure*}[t]
    \centering
    {\epsfig{file=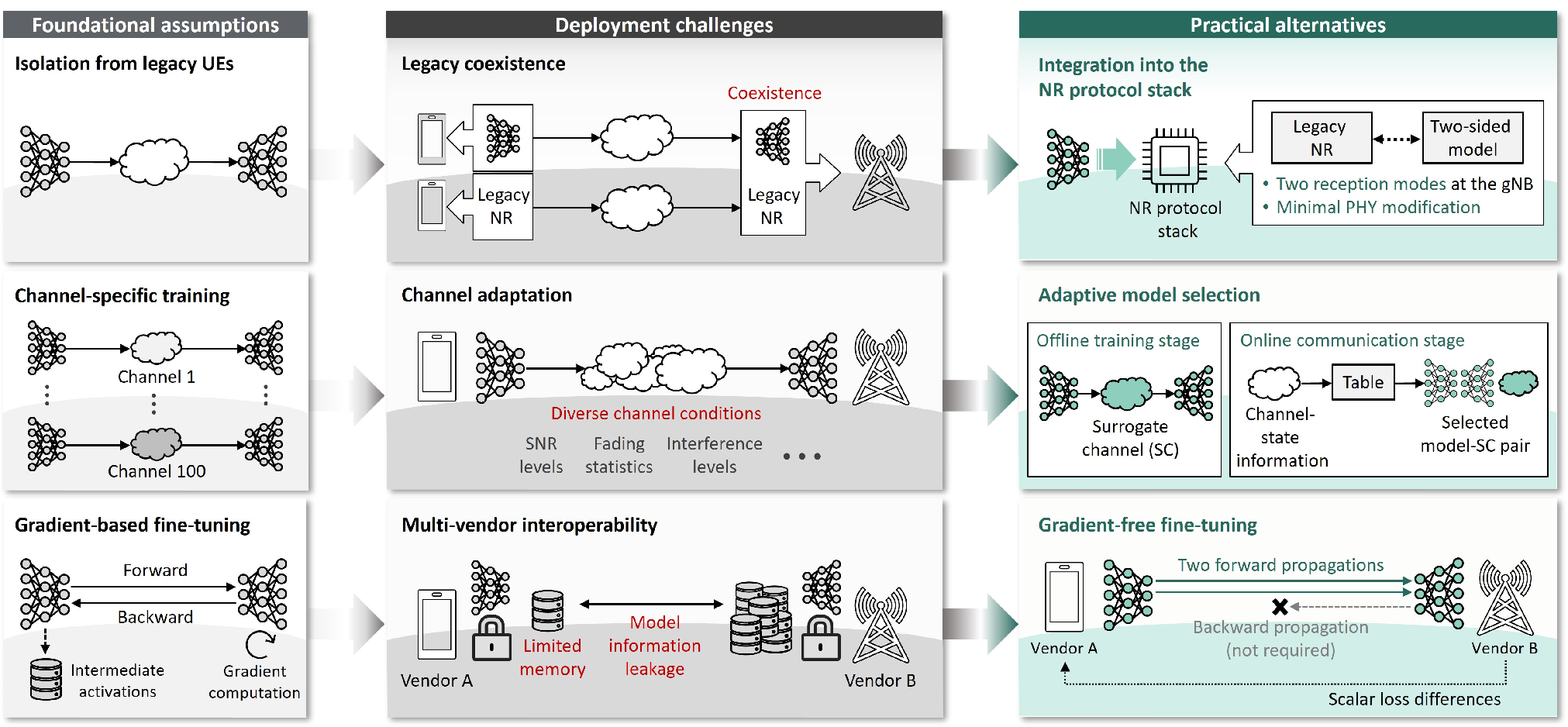, width=18cm}}\vspace{-2mm}
    \caption{Overview of the foundational assumptions, deployment challenges, and corresponding practical alternatives for legacy coexistence, channel adaptation, and multi-vendor interoperability.}\label{fig:overview}\vspace{-3mm}
\end{figure*}

\section{From Foundational Assumptions to Practical Limitations}\label{Sec:Question}

In this section, we examine the common foundational assumptions underlying two-sided models and their limitations in practical deployments, with respect to legacy coexistence, channel adaptation, and multi-vendor interoperability.





\vspace{1mm}
{\bf Legacy Coexistence:} 
Two-sided models are often studied under the assumption that all participating UEs support the same AI-native air interface. 
This assumption conflicts with the emerging 6G deployment direction, in which AI-native 6G operation must coexist with legacy 5G NR, for example through Multi-RAT Spectrum Sharing (MRSS) \cite{ngmn2026architecture}. Moving beyond this assumption therefore requires validating two-sided models in a practical coexistence scenario, where two-sided and conventional NR UEs operate simultaneously. Such validation should demonstrate full-stack compatibility, from physical-layer (PHY) processing to higher-layer procedures, while preserving the performance advantages of two-sided models under coexistence.

\begin{figure*}[t]
    \centering
    {\epsfig{file=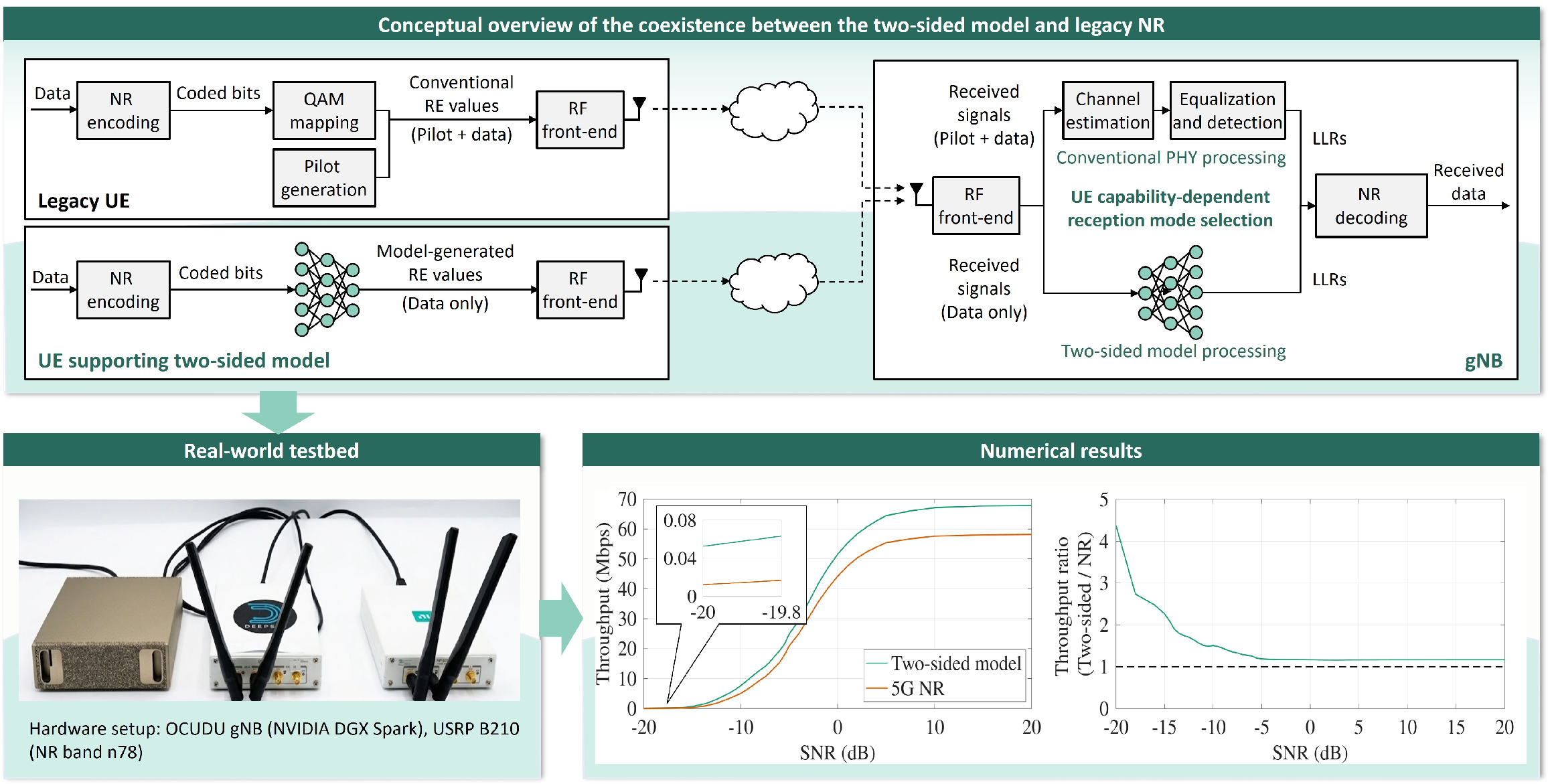, width=17.5cm}}\vspace{-2mm}
    \caption{Integration of the two-sided model into the NR protocol stack, demonstrated through a real-world testbed, along with its performance evaluation for uplink data communication.}\label{fig:arc}\vspace{-3mm}
\end{figure*}

\vspace{1mm}
{\bf Channel Adaptation:} 
For two-sided models, adapting to varying channel conditions is a key aspect of AI model life cycle management (LCM) \cite{3gpp_tr38843_rel19}. Existing methods typically realize such adaptation by incorporating channel conditions into model training \cite{DeepJSCC_Q,10015684}. 
One representative approach is \emph{one-model-fits-one}, where a separate model is trained for each channel condition. Multiple models are prepared in advance, and adaptation is performed by selecting the model best matched to the current channel condition \cite{DeepJSCC_Q}. This approach provides strong channel-specific performance but requires extensive training and model storage to cover diverse channel conditions, limiting its practicality for resource-constrained devices. Another approach is \emph{one-model-fits-all}, where a single model is jointly trained across multiple channel conditions, such as different signal-to-noise ratios (SNRs) \cite{10015684}. Although this approach reduces storage overhead by using a single generalized model, it requires more extensive training and typically sacrifices performance compared with channel-specific models. Both approaches therefore entail substantial deployment tradeoffs, and neither offers a practical solution for channel adaptation.

\vspace{1mm}
{\bf Multi-Vendor Interoperability:} In practical deployments, the transmitter- and receiver-side models are not necessarily deployed by the same vendor, requiring joint training or fine-tuning across vendors. Existing two-sided models typically rely on gradient-based joint training, which 3GPP classifies into Type~1 and Type~2 \cite{3gpp_tr38843_rel19}\footnotemark. In Type~1, both models are trained at either the UE or the gNB, requiring one model to be transferred. In Type~2, each model remains at its respective side, while intermediate model features and gradients are exchanged. Unfortunately, both approaches become impractical when vendors are unwilling to disclose proprietary model information. Type~1 exposes the model architecture and parameters, whereas Type~2 allows exchanged features and gradients to be exploited for model output inference and adversarial attacks \cite{li2024emgan,papernot2016limitations}. Moreover, exchanging models, features, or gradients incurs substantial communication overhead, while Type 2 also requires transmitter-side intermediate activations to be retained until the corresponding gradients are received, imposing a significant memory burden on resource-constrained devices. 

\footnotetext{3GPP also defines Type 3, which alternates local training between the gNB and the UE instead of joint training, but incurs substantial performance degradation \cite{3gpp_tr38843_rel19}.}

\vspace{1mm}
Motivated by the deployment challenges associated with these foundational assumptions of two-sided models, as summarized in Fig.~\ref{fig:overview}, we propose practical alternatives in the following sections.




\section{Integration into the NR Protocol Stack for Legacy Coexistence}
Following the emerging 6G deployment direction \cite{ngmn2026architecture}, we need to consider how two-sided models can be introduced while preserving conventional NR operation. To this end, we propose a capability-dependent PHY integration architecture that introduces two-sided model processing as an alternative to selected conventional NR PHY-functions, preserving its performance advantage while retaining the conventional NR processing path without modification.

To implement and validate this architecture, we integrate two-sided model processing into a real-world transceiver using a full-stack testbed based on DeepSig OmniPHY Axon, as illustrated in Fig.~\ref{fig:arc}. The testbed comprises an open central unit/distributed unit (OCUDU)-based gNB \cite{ocudu2025} running on an NVIDIA DGX Spark and a software-defined UE. The radio-frequency front-end uses a USRP B210 operating in NR band n78 with a 20 MHz bandwidth, 30 kHz subcarrier spacing, and split 8. The same OCUDU-based gNB also supports a WNC open radio unit over an O-RAN 7.2 fronthaul interface with bandwidths of up to 100 MHz. 


The implemented gNB supports two reception modes and selects the appropriate processing path according to the UE's two-sided model capability. In the conventional mode, legacy UEs are served using the existing NR processing chain. In the two-sided model mode, the UE-side model maps coded bits directly to complex-valued symbols without dedicated pilots, replacing conventional quadrature amplitude modulation (QAM) mapping and pilot insertion while retaining NR resource-element (RE) mapping. At the gNB, the corresponding receiver model jointly performs channel estimation and demapping, directly producing soft-bit log-likelihood ratios (LLRs). These LLRs are passed to the existing NR channel decoder, thereby enabling two-sided models for data communication with minimal PHY modifications while maintaining the remaining NR processing stack, including channel decoding.

Importantly, the integration extends beyond PHY processing. The testbed supports UE registration, protocol data unit session establishment, and real-time Internet Protocol traffic, including ping, secure shell, and throughput measurements, with round-trip times of $10$--$20$ ms. These experiments demonstrate full-stack operation of the two-sided model within an operational NR system while preserving conventional NR processing for legacy UEs.

Fig.~\ref{fig:arc} illustrates the overall testbed architecture together with numerical results for uplink data transmission. In this experiment, we consider a 3GPP TR 38.901 rural macro channel with a $1\times 4$ antenna configuration \cite{3gpp38901}. The conventional NR baseline uses pilot-aided linear minimum mean squared error (LMMSE)-Wiener channel estimation and an LMMSE equalizer. The numerical results in Fig.~\ref{fig:arc} compare the uplink throughput of conventional NR and the two-sided model under different SNRs. Both throughput curves are obtained from a single gNB, demonstrating coexistence between the two-sided model and conventional NR operation for legacy UEs. Under this coexistence, the two-sided model achieves more than $4$ times higher throughput than conventional NR under low-SNR conditions.




\begin{figure*}[t]
    \centering
    {\epsfig{file=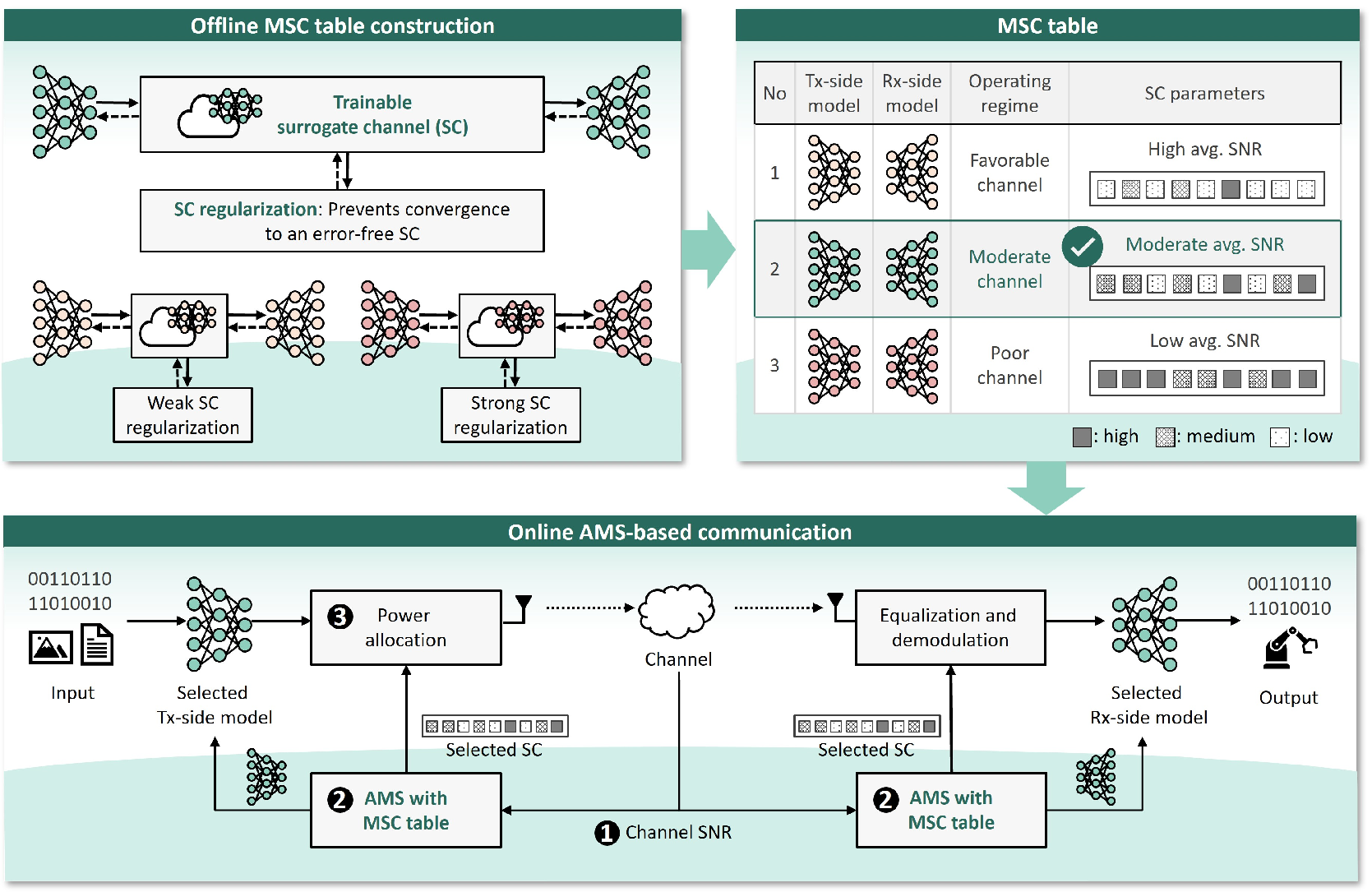, width=14cm}}\vspace{-2mm}
    \caption{The AMS framework for channel adaptation, along with the MSC table construction based on joint optimization of the two-sided model and the SC.}\label{fig:FCQ}
\end{figure*}

\begin{figure*}[t]
    \centering
    {\epsfig{file=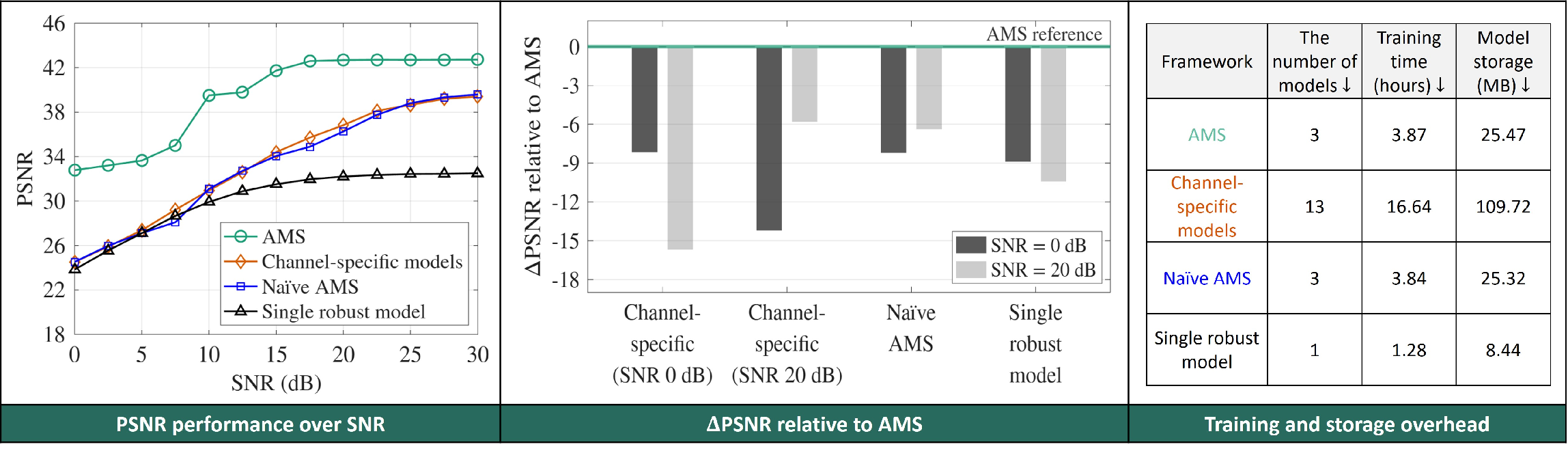, width=17cm}}\vspace{-2mm}
    \caption{Comparison of PSNR, training time, and model storage between the AMS framework and other two-sided model schemes for an image reconstruction task.}\label{fig:Simul_1}\vspace{-3mm}
\end{figure*}

\section{Adaptive Model Selection for Channel Adaptation}

\begin{figure*}[t]
    \centering
    {\epsfig{file=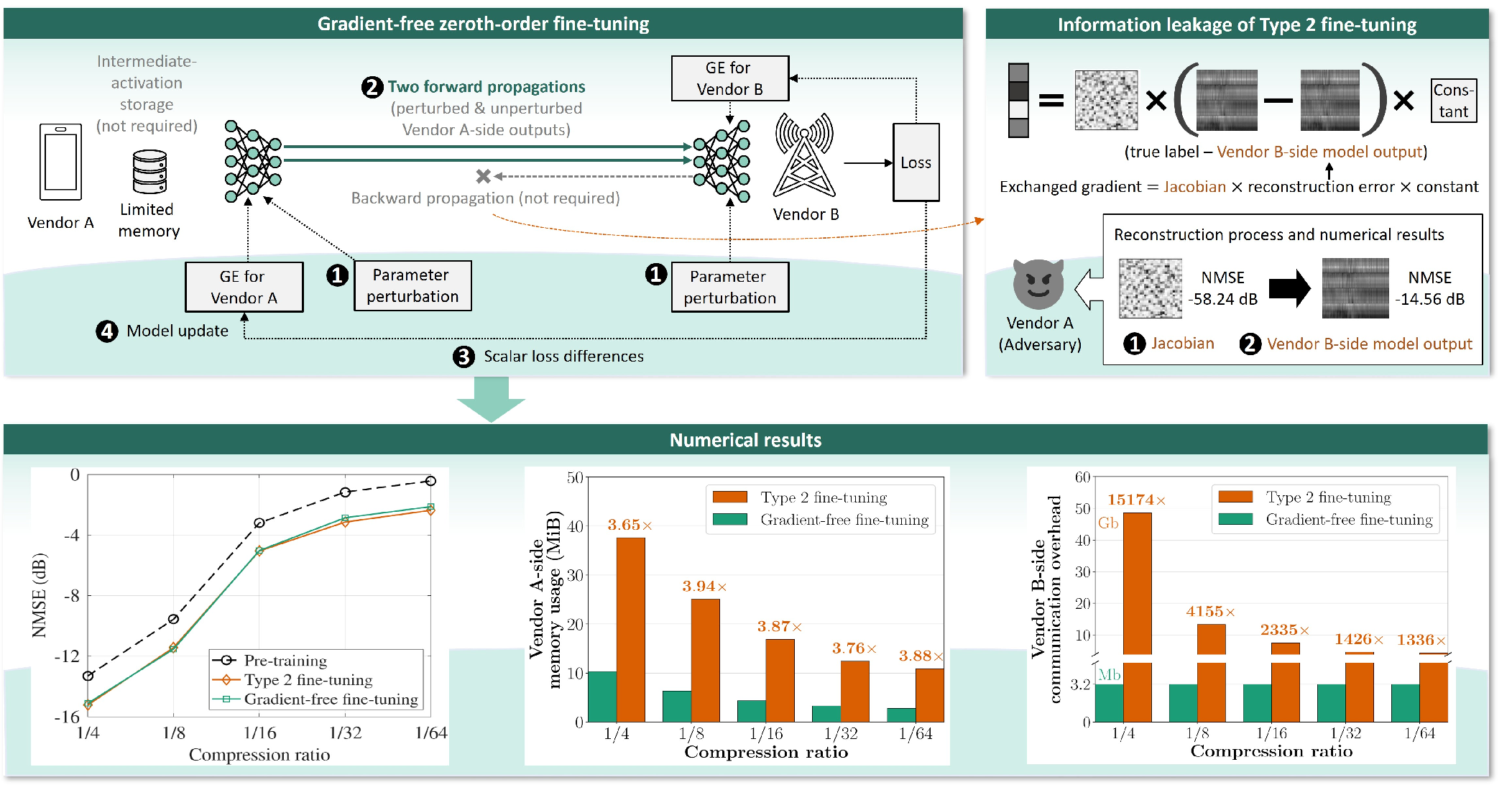, width=18cm}}\vspace{-2mm}
    \caption{The gradient-free zeroth-order fine-tuning framework for multi-vendor interoperability and its performance evaluation for CSI feedback.}\label{fig:zoo}\vspace{-3mm}
\end{figure*}

Channel adaptation is essential for reliable and efficient communication under varying channel conditions. While conventional systems can readily adapt through table-based mechanisms, such as modulation and coding scheme (MCS) tables in adaptive modulation and coding (AMC) and precoding codebooks in beamforming, two-sided models are difficult to adapt rapidly without substantial storage, fine-tuning overhead, or performance loss. To retain the short adaptation delay and low computational overhead of table-based operation, we introduce an adaptive model selection (AMS) framework. AMS selects the model best matched to the current channel condition from an offline-constructed table containing a small number of models together with their associated transmission parameters. Unlike existing one-model-fits-one approaches that pre-train a separate model for each predefined channel condition \cite{DeepJSCC_Q,10015684}, AMS replaces predefined channels with surrogate channels (SCs). These SCs are jointly trained with the two-sided models, followed by instantly mapping into the actual channel condition during operation. The number of SCs is much smaller than the number of possible channel conditions, enabling scalable adaptation across diverse channel conditions, as investigated in \cite{oh2026sfc,oh2025blindtraining}.

Specifically, for two-sided models with analog channel inputs, the SC is modeled as parallel additive white Gaussian noise channels with trainable noise variances \cite{oh2026sfc}. For digital transmission, it is modeled as parallel binary symmetric channels with trainable bit-flip probabilities \cite{oh2025blindtraining}. The trainable SC parameters and the two-sided model are jointly optimized to minimize a weighted sum of the task loss and an SC regularization term. Without 
regularization, the SC converges to a trivial error-free channel because lower distortion improves task performance. The regularization prevents this by penalizing small SC parameters and encouraging non-zero channel distortion. Each jointly optimized model-SC pair forms an entry in the model-and-SC (MSC) table. Varying the regularization level yields entries with different average channel distortion levels.

During online operation, each trained SC is reproduced over the actual wireless channel by adjusting the transmission parameters. For analog transmission, the trained noise variance is converted to a target SNR, which is matched through transmit power control \cite{oh2026sfc}. For digital transmission, the trained bit-flip probability is treated as a target bit error rate and matched through transmit power and/or modulation control \cite{oh2025blindtraining}. Under the current channel condition, AMS determines the transmission parameters required for each MSC entry and identifies those satisfying the available resource constraints, such as total transmit power. Among the feasible entries, it selects the one achieving the highest task performance.

Fig.~\ref{fig:Simul_1} presents numerical results for an image reconstruction task on the CIFAR-$10$ dataset. The channel follows Rayleigh fading over 64 subcarriers with $16$-QAM. Training time is measured on an Intel Core i$7$-$12700$ CPU, an NVIDIA RTX $3090$ GPU, and $32$ GB RAM. Three baseline schemes are used for comparison. The one-model-fits-one baseline trains a separate channel-specific model for each SNR, whereas the one-model-fits-all baseline trains a single robust model over the $0$--$30$ dB SNR range. Na\"ive AMS uses three models trained over the $0$--$10$, $10$--$20$, and $20$--$30$ dB SNR ranges. Task performance is measured by the peak SNR (PSNR) of the reconstructed image at the receiver.

The results show that AMS achieves the highest PSNR across the entire $0$--$30$ dB SNR range. The one-model-fits-one baseline achieves the second-highest PSNR, but incurs significant training and storage overhead, with severe performance degradation outside the SNRs used for training. Although the one-model-fits-all baseline has the lowest overhead, it also yields the lowest PSNR over most of the SNR range. Na\"ive AMS has overhead comparable to AMS, but its PSNR is $6$--$8$ dB lower over the $0$--$20$ dB range, underscoring the importance of MSC construction and transmission parameter selection.

\section{Gradient-Free Fine-Tuning for Multi-Vendor Interoperability}
Multi-vendor interoperability of two-sided models requires joint fine-tuning of the transmitter- and receiver-side models held by different vendors. The current 3GPP Type~1 and Type~2 fine-tuning approaches \cite{3gpp_tr38843_rel19} require the exchange of model parameters and gradients, respectively, to perform backpropagation, which may raise privacy concerns and incur additional memory and communication overhead. To overcome this limitation, we present a gradient-free fine-tuning framework that requires only scalar feedback while reusing the existing communication path for forward propagation. The framework is grounded in the zeroth-order optimization method \cite{chen2025zeroth}, a gradient-free approach that estimates a descent direction using only forward evaluations of the loss. 

To illustrate, suppose a scenario in which Vendor ${\sf A}$ and Vendor ${\sf B}$ hold the transmitter-side and receiver-side models, respectively. Vendor ${\sf A}$ perturbs its model parameters and performs two forward propagations with the original and perturbed parameters, transmitting both outputs to Vendor ${\sf B}$. Using these outputs and ground-truth labels from a shared fine-tuning dataset, Vendor ${\sf B}$ evaluates the loss with its original and perturbed model parameters. The resulting loss difference, together with the corresponding parameter perturbation, provides a zeroth-order estimate of the gradient direction. Vendor ${\sf B}$ uses this estimate to update its model and returns the scalar loss difference to Vendor ${\sf A}$, which similarly estimates the gradient direction and updates its model. The procedure naturally extends to multiple random perturbations, whose gradient estimates are averaged to obtain a more reliable batch gradient. The corresponding perturbed outputs can be transmitted together in a single batch, allowing multiple perturbations without additional communication rounds.

Unlike Type 1 fine-tuning, which requires model parameter exchange, and Type 2 fine-tuning, which requires both forward and backward propagations between Vendors ${\sf A}$ and ${\sf B}$, the proposed method relies only on forward propagations, with scalar loss differences fed back to Vendor ${\sf A}$. This limited feedback reduces the risk of exposing proprietary information about the Vendor ${\sf B}$-side model, as gradient feedback may enable Vendor ${\sf B}$-side model output inference and adversarial attacks by Vendor ${\sf A}$ \cite{papernot2016limitations}. Furthermore, the proposed method substantially reduces Vendor ${\sf B}$-side communication overhead, as the scalar feedback is small and its size is independent of the Vendor ${\sf A}$-side model output dimension. Finally, eliminating gradient calculation removes the need to retain intermediate activations generated during forward propagation until the subsequent backward propagation, thereby reducing the memory burden on resource-limited devices.

Fig.~\ref{fig:zoo} illustrates the overall pipeline of the gradient-free zeroth-order fine-tuning framework and presents numerical results. In the simulation, we consider a two-sided model for CSI feedback in an urban macro cell scenario, where the CSI is represented as a $32 \times 32$ angular-delay matrix. The carrier frequency is set to $3.5$~GHz, and a uniform linear array with half-wavelength antenna spacing is used. 
The two-sided model is first pre-trained in this scenario and then fine-tuned for a different cell geometry. The loss function is the normalized mean squared error (NMSE). Using this setup, we evaluate the CSI reconstruction performance, memory usage, communication overhead, and information leakage associated with gradient exchange in conventional Type 2 fine-tuning. 

As illustrated in the upper-right part of Fig.~\ref{fig:zoo}, the exchanged gradient can be expressed as the product of the Jacobian matrix, the reconstruction error, and a scaling constant by applying the chain rule, where the Jacobian is defined as the derivative of the Vendor ${\sf B}$-side model output with respect to the Vendor ${\sf A}$-side output. Access to this Jacobian reveals how the Vendor ${\sf A}$-side output influences the Vendor ${\sf B}$-side model output, potentially allowing an adversary to craft perturbations that manipulate the latter \cite{papernot2016limitations}. In Type 2 fine-tuning, an honest-but-curious Vendor ${\sf A}$ can estimate this Jacobian from the exchanged gradients. 
Specifically, Vendor ${\sf A}$ deliberately transmits the same model output for different CSI samples and subtracts the corresponding returned gradients, yielding the product of the transposed Jacobian and the difference between the CSI samples. By repeating this process with sufficiently diverse CSI samples and jointly solving the resulting linear equations, Vendor ${\sf A}$ can recover the full Jacobian matrix. The recovered Jacobian can further be used to estimate the Vendor ${\sf B}$-side model output from the observed gradients. 

%

At a compression ratio of $1/4$, simulation results show that Vendor ${\sf A}$ can accurately estimate both the Jacobian matrix and the Vendor ${\sf B}$-side model output from the gradients exchanged during Type 2 fine-tuning, achieving NMSE values of $-58.24$ dB and $-14.56$ dB, respectively. In contrast, the proposed gradient-free fine-tuning framework eliminates gradient feedback, thereby preventing such information leakage. The bottom part of Fig.~\ref{fig:zoo} further shows that gradient-free fine-tuning achieves CSI reconstruction NMSE comparable to that of conventional Type 2 fine-tuning, while reducing Vendor ${\sf A}$-side memory usage by $3.65$ times and Vendor ${\sf B}$-side communication overhead by more than $15,000$ times. These gains are consistently achieved across different compression ratios.

\section{Open Challenges and Future Directions}
The preceding sections have revisited the foundational assumptions of two-sided models and introduced practical alternatives for legacy coexistence, channel adaptation, and multi-vendor interoperability. These advances motivate new research directions while highlighting further challenges that must be addressed for the  practical deployment of two-sided models, as discussed next.

\vspace{1mm}
{\bf AMC-AMS Unification:} Conventional AMC uses channel quality indicator (CQI) feedback to select a modulation level and code rate from the MCS table. AMS can similarly use CQI feedback to select a model-SC pair from the MSC table. This suggests a unified adaptation framework that jointly coordinates AMC and AMS for a given CQI feedback. Validating such joint AMC-AMS operation while guaranteeing legacy coexistence remains an important challenge.


\vspace{1mm}
{\bf Extension to Task-Oriented Communication:} 
Extending two-sided models from CSI feedback and conventional data communication to task-oriented communication is promising but non-trivial. In task-oriented communication, model input and output data reside at the application layer, while channel information remains in the radio access network (RAN), requiring costly cross-layer exchange. Joint model-SC training avoids this during training, but online operation still requires visibility into application-layer task-performance degradation and its causes. This calls for standardized cross-layer interfaces or co-located PHY and application-layer processing in AI-RAN \cite{airan2026architecture}.

\vspace{1mm}
{\bf Multi-Vendor LCM:} In multi-vendor scenarios, the MSC table can be split into transmitter- and receiver-side tables, allowing each vendor to maintain its own model while sharing only the SCs. This raises multi-vendor LCM issues, including who determines the number of table entries and how they are updated over time. Similar questions arise for fine-tuning, such as when it should be triggered and which vendor should initiate it. Standardized or vendor-agreed rules are therefore required.

\section{Conclusion}
This article has examined key deployment challenges of two-sided AI models, in terms of legacy coexistence, channel adaptation, and multi-vendor interoperability. To address these challenges, we presented three practical alternatives: NR protocol-stack integration, MSC-table-based AMS, and gradient-free zeroth-order fine-tuning. We also discussed open challenges, such as the unification of AMS and AMC, extensions to task-oriented communication, and multi-vendor LCM. Addressing these challenges through coordinated advances in research, prototyping, and standardization will be critical to moving two-sided models beyond isolated demonstrations towards interoperable and deployable components of next-generation air interfaces.





\bibliographystyle{IEEEtran}
\bibliography{Reference}

\end{document}